
\magnification=\magstep1
\hsize 6.0 true in
\vsize 9.0 true in

\voffset = - .2 true in

\font\tentworm=cmr10 scaled \magstep2
\font\tentwobf=cmbx10 scaled \magstep2

\font\tenonerm=cmr10 scaled \magstep1
\font\tenonebf=cmbx10 scaled \magstep1

\font\eightrm=cmr8
\font\eightit=cmti8
\font\eightbf=cmbx8
\font\eightsl=cmsl8
\font\sevensy=cmsy7
\font\sevenm=cmmi7

\font\twelverm=cmr12
\font\twelvebf=cmbx12
\def\subsection #1\par{\noindent {\bf #1} \noindent \rm}

\def\mid {\let\rm=\tenonerm \let\bf=\tenonebf \rm \bf}

\def\para{\par \vskip 12 pt}

\def\head{\let\rm=\tentworm \let\bf=\tentwobf \rm \bf}

\def\heading #1 #2\par{\centerline {\head #1} \smallskip
 \centerline {\head #2} \vskip .15 pt \rm}

\def\eight{\let\rm=\eightrm \let\it=\eightit \let\bf=\eightbf
\let\sl=\eightsl \let\sy=\sevensy \let\m=\sevenm \rm}

\def\foots{\noindent \eight \baselineskip=10 true pt \noindent \rm}
\def\sexion{\let\rm=\twelverm \let\bf=\twelvebf \rm \bf}

\def\section #1 #2\par{\vskip .35true in \noindent {\mid #1} \enspace {\mid #2}
  \para \noindent \rm}

\def\abstract#1\par{\para \foots {\bf Abstract: \enspace}#1 \para}

\def\author#1\par{\centerline {#1} \vskip 0.1 true in \rm}

\def\abstract#1\par{\noindent {\bf Abstract: }#1 \vskip 0.5 true in \rm}

\def\midsection #1\par{\noindent {\sexion #1} \noindent \rm}

\def\sqr#1#2{{\vcenter{\vbox{\hrule height.#2pt
  \hbox {\vrule width.#2pt height#1pt \kern#1pt
  \vrule width.#2pt}
  \hrule height.#2pt}}}}

\baselineskip=18pt
\pretolerance=10000

\def\m{\medskip}

\def\n{\noindent}


\centerline{\mid TOWARDS A PATH INTEGRAL FOR PURE SPIN CONNECTION}

\smallskip
\centerline{\mid FORMULATION OF GRAVITY}

\vskip 1.50 true in

\centerline{Abhijit K. Kshirsagar \footnote{$\dagger$}{e-mail address~:
abhi@iucaa.ernet.in}
}
\centerline{Inter-University Centre for Astronomy and Astrophysics}
\centerline{Post Bag 4, Ganeshkhind, Pune 411 007}
\centerline{INDIA}

\vskip 2.0 true in
\centerline{\bf Abstract}

\medskip
A proposal for the path-integral of pure-spin-connection formulation of gravity
is
described, based on the two-form formulation of Capovilla et. al. It is shown
that the resulting effective-action for the spin-connection, upon functional
integration of the two-form field $\Sigma$ and the auxiliary matrix field
$\psi$
is {\it non-polynomial}, even for the case of vanishing cosmological constant
and absence of any matter couplings. Further, a diagramatic evaluation is
proposed
for the contribution of the matrix-field to the pure spin connection action.

\bigskip\bigskip\noindent
{\bf PACS} numbers :-  04.20.Cv, 04.20.Fy, 2.40.+m

\vfill\eject


In the past few years, it has become increasingly evident that the description
of gravity in terms of connection variables instead of the metric, originally
due to Ashtekar [1] is well-tailored for the discussion of quantum aspects
of the theory. This has been attributed to the close parallel between this
description
and Yang-Mills theories, topological solutions thereof as well as the invention
of loop variables.

Motivation to obtain a natural covariantization of Ashtekar theory, led
Capovilla et al to introduce a classical action for gravity (and a
one-parameter
family of generally covariant gauge-theories) purely in terms of a
spin-connection
[2]. This action is obtained by solving the classical equation of motion for
the `metric-variable' $\Sigma$, from the self-dual two form action for a
$SL(2,C)$ connection
$A$, a non-dynamical matrix field $\psi$ and two-forms $\Sigma$ [3]. The
equivalence
of the pure-connection theory to that of Ashtekar can be shown by a $3+1$
decomposition [4], as well as by comparing the constraints arising due to
diffeomorphism
and gauge invariances of the theory in the two formulations [5]. In an
interesting
alternative approach, Peld\'an [6] performed inverse Legendre transform on the
Hamiltonian
comprising purely of constraints (characteristic of diffeomorphism invariant
theories)
and obtained a pure-spin connection action. The apparent discrepancy between
the actions of Peld\'an and Capovilla et al, can be removed by rewriting the
tracelessness condition on $\psi$ in the pure-spin connection action [7].

With an overall agreement on the consistency of the pure-spin-connection
formulation
of gravity at the classical level at hand, it is only natural now to start
exploring
the quantum properties of it. In a recent paper, Smolin [8] has furnished a
path
integral for Euclidean case, starting from the Hamiltonian of the `googly'
theory.
For eliminating the Gauss law and diffeomorphism constraints from the
integrand, he
uses the `time' component of the gauge field as a Lagrange multiplier and
solves
the diffeomorphism constraints explicitly using the {\it classical} solution of
Capovilla et al [2]. He further proposes to choose gauge-fixing conditions for
the
$A$ field as linear expressions, so that the gauge field action remains at most
quadratic
(this happens only in the limit $G_N \rightarrow 0$, and hence for the `googly'
theory alone) and the path integral can, be evaluated exactly, producing an
`effective
action' for the matrix field and the ghost fields introduced by the
Faddeev-Popov
determinant. Although motivated in part by Smolin's paper, we wish to make
a different proposal for the path integral. We begin with the two-form action
for
the metric variable $\Sigma$, coupled to the gauge-field $A$ and the auxiliary
matrix
field $\psi$, as in reference [3]. As in any quantum theory of gravity, the
path
integral must include fluctuations of the metric, we functionally integrate
over $\Sigma$
first to obtain the effective action for the gauge-field $A$ and $\psi$.
Throughout
the discussion the integral over $A$ is only in a formal sense, since we
do not display the gauge-fixing terms and the $F-P$ determinant, those
will be discussed in a future publication as work is still in progress on these
issues.

Consider then the following formal definition of the Euclidean path integral~:

$$Z = \int DA D \psi D \Sigma~ e^{-\int \Sigma^a \wedge F^a + {1\over 2}
\int \psi_{ab} \Sigma^a \wedge \Sigma^b - S_{gf} - S_{FP}} \delta (tr \psi)
\eqno(1)$$

\n where $S_{gf}$ and $S_{FP}$ are the gauge-fixing and Faddeev-Popov terms in
the
action needed for the path integral over the gauge field $A_\mu^a$. The matrix
$\psi_{ab}$ is symmetric and $\delta (tr \psi)$ denotes the constraint in the
path
integral that $\psi$ should be traceless.

To perform the path integral over $\Sigma$, we follow the standard procedure
and write

$$\Sigma^a = \bar \Sigma^a + \sigma^a \eqno(2)$$

\n where $\sigma^a$ is the fluctuation part and $\bar \Sigma^a$ satisfies the
classical
equation of motion

$$F_a = \psi_{ab} \Sigma^b \eqno(3)$$

Substituting in the path integral we get

$$Z = \int DAD\psi D\sigma~ e^{-S} . \delta (tr \psi )\eqno(4)$$

\n where the action $S$ is given by

$$S = {1\over 2}\int \psi^{-1 ab} F_a \wedge F_b - {1\over 2}\int
\psi_{ab} \sigma^a \wedge \sigma^b + S_{gf} + S_{FP}\eqno(5)$$

\n The integration over $\sigma^a$ then produces the factor ${1\over(det
{}~\psi)}$
in the path integral (in Euclidean signature, $\Sigma$'s are hermitian) and we
write

$$Z = \int DAD \psi e^{-\int {1\over 2} \psi^{-1 ab} F_a \wedge F_b -
S_{gf} - S_{FP}} {1 \over (det~\psi)} . \delta (tr \psi )\eqno(6)$$

\n In order to perform next the integration over the auxiliary field $\psi$,
it is convenient to perform a change of variables from $\psi \rightarrow
\phi = \psi^{-1}$. (Invertibility of $\psi$ is anyway assumed in the pure
spin connection formulation and thus existence of this transformation
is no additional assumption.) The functional integration measure changes
accordingly as

$$D\psi \rightarrow D \phi \bigg\vert det~{\partial \psi \over \partial \phi}
\bigg\vert$$

\n where the jacobian of the transformation can easily be seen to be

\noindent

\centerline{$det (\psi^2) = (det \psi)^2 = {1\over (det \phi)^2}$.}

Rewriting the reciprocal determinant arising from $\Sigma$ integral in terms of
$det
\phi$, we get

$$Z = \int DAD\phi {1 \over (det ~\phi)} e^{-\int {1\over 2} \phi^{ab}
F_a \wedge F_b - S_{gf} - S_{FP}} . \delta (tr \psi )\eqno(7)$$

The constraint that $\psi$ is traceless is equivalent to the constraint on its
inverse,
viz.

$$(tr \phi )^2 - tr \phi^2 = 0 \eqno(8)$$

\n which follows from the characteristic equation satisfied by a non-degenerate
$3 \times
3$ matrix. We use this equivalent form also because it is this form which leads
to the agreement between the actions of Capovilla et al and Peld\'an [5,7]. The
delta
function imposing this constraint can be promoted to the action by using its
functional
representation via the introduction of a (complex) auxiliary field $\mu$ [9],
as

$$Z = \int DAD \phi D\mu e^{-\int {1\over 2} \phi^{ab} F_a \wedge F_b -
\int \mu [ (tr \phi)^2 - tr \phi^2 ] - S_{gf} - S_{FP}} . {1 \over det~\phi}
\eqno(9)$$

The inverse determinant factor can also be promoted to the action as

$${1\over det~\phi} = e^{- \delta^{(4)} (0) \int tr~ln \phi}\eqno(10)$$

\n where the zero-momentum delta function is just the volume of (Euclidean)
four space and is to be understood in a regularized sense. We suppress
this factor henceforth [10].

The path integral can then finally be written over the gauge field, $\psi$
and $\mu$ of the `effective-action'

$$S_{eff} = \int {1\over 2} \phi^{ab} F_a \wedge F_b + \int \mu
[(tr \phi)^2 - tr \phi^2 ] + tr~ln \phi + S_{gf} + S_{FP} \eqno(11)$$

Several remarks are in order at this point. In order to compute the
`effective-action' for the spin-connection, one now needs to integrate over
$\phi$ and $\mu$ besides the ghost fields. The action for $\phi$ is
however no more quadratic and so the integral can at best be evaluated by
methods of standard perturbation theory. Even treating this as
the {\it classical action} $S_{eff} [A, \phi , \mu]$ for $\phi$, the equation
of motion

$${1\over 2}\epsilon^{\mu\nu\rho\lambda} F_{a\mu\nu} F_{b\rho\lambda} + 2 \mu
[(tr \phi) \delta_{ab} - \phi_{ab} ]
+ \phi^{-1}_{ab} = 0 \eqno(12)$$

\n can not be solved in a closed form for $\phi$ in terms of the curvature of
the
spin-connection. The upshot of all this is that even for the vanishing
cosmological
constant case, the $S_{eff} [A, \phi , \mu]$ is {\it non-polynomial} in $F$
as opposed to the quartic action obtained by solving the equation of motion for
the
two-form $\Sigma$ $\acute {\rm a}$ la Capovilla et al in [3]. Thus the effect
of
including the fluctuations in $\Sigma$ is to render the spin-connection action
non-polynomial in curvature.

So, to obtain the effective action for the spin-connection we now set up the
perturbation theory and obtain the associated Feynman rules for the field
$\phi$.
For this purpose it is convenient to make one more change of variables,
viz. $\phi \rightarrow \tilde \phi = \phi -1$. The path integral then becomes

$$Z = \int DAD \tilde \phi D \mu e^{-S[A, \tilde \phi ]}\eqno(13)$$

\m where

$$\eqalignno{S[A, \tilde \phi] = \int {1\over 2} \tilde \phi^{ab} F_a \wedge
F_b &+ \int {1\over 2} tr F \wedge F + \int tr ~ln (1 + \tilde \phi)\cr
&- \int {1\over 2} \mu [(tr \tilde \phi)^2 - tr \tilde \phi^2 ] -
\int \mu [3 + 2 tr \tilde \phi]&(14)\cr}$$

It is interesting to note the presence of the toplogical term $tr F \wedge F$
in this
action. We can now read off the Feynman rules from the $\phi$-part of the
action~:

$$S[\tilde \phi] = {1\over 2} \int \tilde \phi^{ab} M_{ab,cd} \tilde \phi^{cd}
+ {1\over 2} \int J_{ab} \tilde \phi^{ab} - \int \sum\limits_n {(-1)^n \over n}
tr \tilde \phi^n \eqno(15)$$

\n where

$$M_{ab,cd} = \mu (\delta_{ab} \delta_{cd} - \delta_{ac}
\delta_{bd})\eqno(16)$$

\n the inverse of which defines the (non-dynamical) `propagator' for the field
$\phi$
and

$$J_{ab} = \epsilon ^{\mu \nu \rho \lambda}F_{a\mu \nu}F_{b\rho \lambda} + 4
\mu \delta_{ab} \eqno(17)$$

\n is the `source' coupling linearly to $\phi$.

The last term implies arbitrary-order self-interactions of the field $\phi$. It
is now
clear that the effective action for the spin-connection will involve arbitrary
powers
of $F \wedge F$, obtained by summation of the diagrams arising from the
$\phi^n$
vertices. In the absence of these vertices, one would obtain precisely the
action
of reference [2] for zero cosmological constant. They produce terms in the
action,
proportional to the zero-momentum delta function.

Work is in progress on evaluating these contributions to the effective action
for the
spin-connection as well as choice of gauge conditions and computation of the
$F-P$
determinant and will be dealt with in a future publication.

To summarize,
we have explicitly demonstrated that the effect of quantum fluctuations of the
`metric-variable' $\Sigma$ is to render the pure-spin connection action non
polynomial.
Needless to say, it remains so even for non-zero cosmological constant as well
as
couplings to other fields. We have further set up Feynman rules for the
diagramatic
evaluation of the effective-action. A byproduct of the perturbation expansion
for $\phi$ is the generation of the topological term for the gauge field
representing
spin-connection.

It is a pleasure to acknowledge useful conversations with Abhay Ashtekar and
Naresh Dadhich and correspondence with the Maryland Relativity Group.

\vfill\eject

\centerline{\mid References}

\bigskip

\item{1.}Ashtekar A., Plenary talk given at the Sixth Marcel Grossman meeting
on
general relativity at Kyoto, June 23-29, 1991. To appear in the proceedings,
World Scientific Publishing Company (1992), ed. T. Nakamura, see also the
references therein.
\smallskip

\item{2.}Capovilla R., Jacobson T. and Dell J. 1989, {\it Phys. Rev. Lett.}
{\bf 63},
2325

\item{}Capovilla R., Dell J. and Jacobson T. 1991, {\it Class. Quant. Gravity}
{\bf 8},
59, see also erratum (to be published).

\smallskip
\item{3.}Capovilla R., Dell J., Jacobson T. and Mason L. 1991 {\it Class.
Quant. Gravity}
{\bf 8}, 41

\smallskip
\item{4.}Capovilla R. 1992 {\it Nucl. Phys.} {\bf B373}, 233

\smallskip
\item{5.}Dadhich N., Koshti S. and Kshirsagar A. 1991 {\it Class. Quant.
Gravity}
{\bf 8}, L61

\item{}Kshirsagar A., contributed talk given at the Sixth Marcel Grossman
meeting on general relativity at Kyoto, June 23-29, 1991. To appear in the
proceedings, World Scientific Publishing Company (1992), ed. T. Nakamura

\smallskip
\item{6.}Peld\'an P. 1990, {\it Phys. Lett.} {\bf B248}, 62

\item{}Peld\'an P. 1991, {\it Class. Quant. Gravity} {\bf 8}, 1765
\smallskip
\item{7.}Capovilla R. and Jacobson T. 1992, University of Maryland preprint
UMDGR-91-134

\smallskip
\item{8.}Smolin L. 1992, {\it Class. Quant. Gravity} {\bf 9}, 883

\smallskip
\item{9.}Zinn-Justin J. 1991, Quantum Field Theory and Critical Phenomena,
Oxford
University Press, p. 97

\smallskip
\item{10.}Faddeev L.D. and Slavnov A.A. 1981, Gauge Fields, Introduction to
Quantum Theory, Addison Wesley Publishing Company, p. 113

\bye